\documentclass[conference]{IEEEtran}
\IEEEoverridecommandlockouts
\usepackage{cite}
\usepackage{amsmath,amssymb,amsfonts}
\usepackage{algorithmic}
\usepackage{graphicx}
\usepackage{textcomp}
\usepackage{xcolor}

\newcommand{\etal}{\textit{et al}. }
\newcommand{\ie}{\textit{i}.\textit{e}.}

\usepackage{geometry}
\def\BibTeX{{\rm B\kern-.05em{\sc i\kern-.025em b}\kern-.08em
    T\kern-.1667em\lower.7ex\hbox{E}\kern-.125emX}}
\begin{document}

\title{High-speed Millimeter-wave 5G/6G Image Transmission via Artificial Intelligence \\
\thanks{*Shaolin Liao is the corresponding author.} }

\author{\IEEEauthorblockN{Shaolin Liao*}
\IEEEauthorblockA{\textit{Department of Electrical and Computer Engineering} \\
\textit{Illinois Institute of Technology}\\
Chicago, IL, USA \\
sliao5@iit.edu; ORCID: 0000-0002-4432-3448}
\and
\IEEEauthorblockN{Lu Ou}
\IEEEauthorblockA{\textit{College of Computer Science and Electronic Engineering} \\
\textit{Hunan University}\\
Changsha, Hunan, China \\
oulu9676@gmail.com} 
}

\maketitle

\begin{abstract}
Artificial Intelligence (AI) has been used to jointly optimize a mmWave Compressed Sensing (CS) for high-speed 5G/6G image transmission. Specifically, we have developed a Dictionary Learning Compressed Sensing neural Network (DL-CSNet) to realize three key functionalities: 1) to learn the dictionary basis of the images for transmission; 2) to optimize the Hadamard measurement matrix; and 3) to reconstruct the lossless images with the learned dictionary basis. A 94-GHz prototype has been built and up to one order of image transmission speed increase has been realized for letters ``A" to ``Z".
\end{abstract}


\section{Introduction}
Millimeter Wave (mmWave) 5G/6G wireless network \cite{han_achieving_2019,  bjornson_massive_2019} has great potential for High-Definition (HD) images transmission and videos streaming \cite{kim_strategic_2017}, due to its broad spectrum (30 GHz -300 GHz). mmWave has been extensively studied for many important applications \cite{liao_sub_thz_2007, Bakhtiari_2011, Liao_2013, liao_novel_2014, liao_four_frequency_2009, liao_fast_2009}.  Also, short-wavelength mmWave band allows massive Multiple Input Multiple Output (MIMO) antenna arrays \cite{bjornson_massive_2019}, which technique explores the spatial diversity of communication channels to dramatically increase the transmission speed.


In addition, Compressed Sensing (CS) is another effective way to increase the data rate by up to 10 folds \cite{Emmanuel2008, s._d._babacan_compressive_2011,  gopalsami_compressive_2011, n._gopalsami_compressive_2011-1, nachappa_gopalsami_passive_2012}. Compared to the software compression, CS measurement matrix operations can be implemented on the hardware level through massive MIMO antenna arrays, which mean faster speed.

What's more, Dictionary Learning (DL) is one such machine learning technique that studies the common dictionary basis of a large amount of images based on their similarities \cite{singh_deep_2017}, which can be used to obtain the optimized CS measurement matrix for even faster speed by minimizing the mutual coherence of the CS measurement matrix\cite{Obermeier-2017}.

All of these call for the joint optimizations of the mmWave image transmission system. In this paper, we apply Artificial Intelligence (AI) and develop the Dictionary Learning Compressed Sensing neural Network (DL-CSNet) for the optimized mmWave 5G/6G image transmission system. Also, a 94-GHz experimental prototype is built to validate the design. 
 
\begin{figure*}[th]
\centerline{\includegraphics[width=0.65\textwidth]{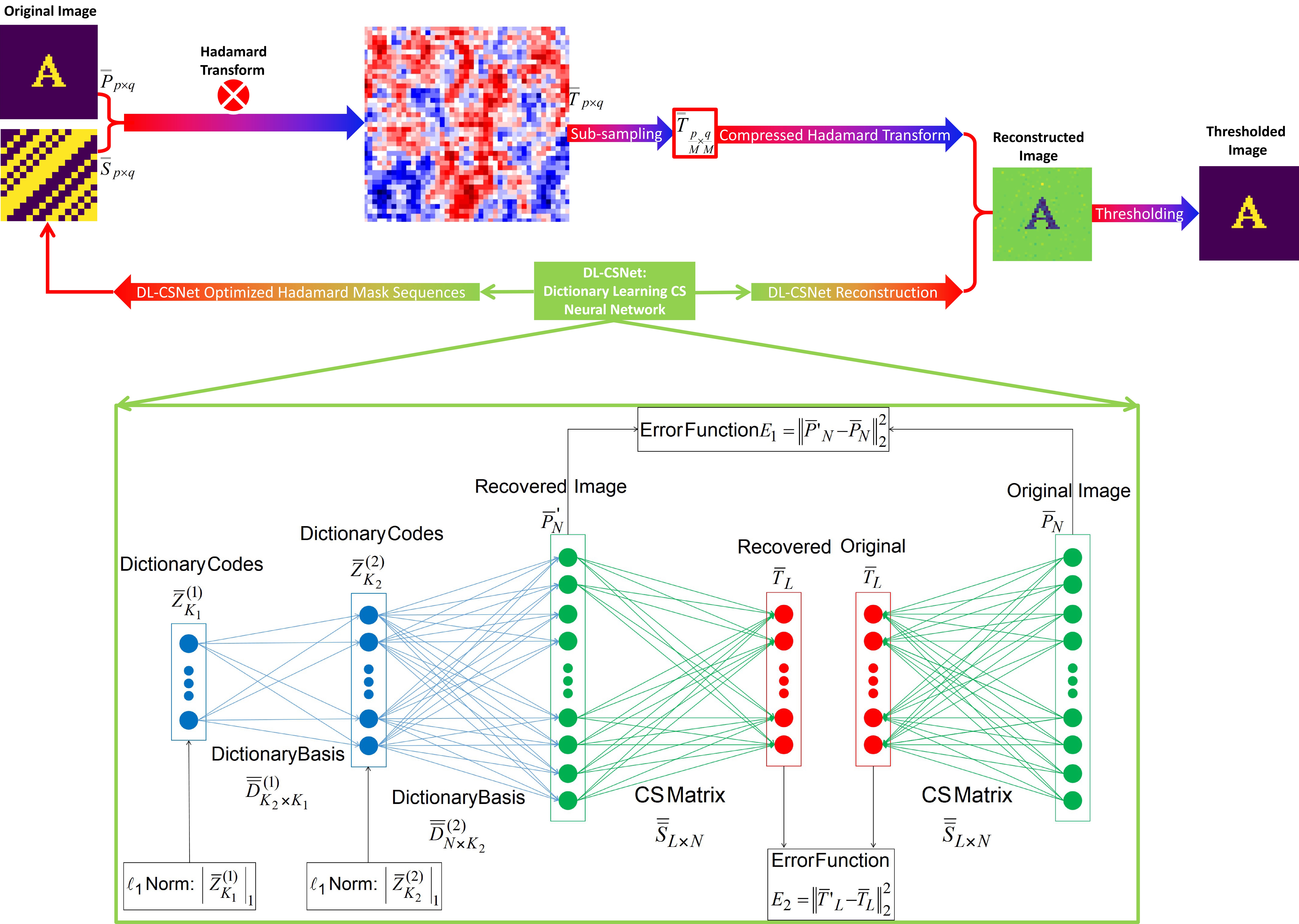}}
\caption{The DL-CSNet optimized mmWave Hadamard-CS iamge transmission.}
\label{fig:flowchart}
\end{figure*} 

\section{The DL-CSNet}

\subsection{The Basic Hadamard-CS Image Transmission} 
 The working principle and the functional blocks of the basic CS image transmission with the Hadamard mask \cite{s._d._babacan_compressive_2011,  gopalsami_compressive_2011, n._gopalsami_compressive_2011-1, nachappa_gopalsami_passive_2012} as the measurement matrix is shown on the top of Fig. \ref{fig:flowchart}:
 
 \subsubsection{On the Transmitter Side}
The original image is first transformed to the CS domain with the Hadamard measurement matrix,
\begin{flalign}\label{eqn:CS_origin}
\overline{\mathcal{T}}_{L} = \overline{\overline{\mathcal{S}}}_{L \times N} {\overline{\mathcal{P}}}_{N} + \overline{n}_L, 
 \end{flalign} 
where $\overline{\mathcal{T}}_{L}$ denotes the 1D vector of the 2D Hadamard-CS image transform of length $L$; $\overline{\overline{\mathcal{S}}}_{L \times N}$ is the Hadamard-CS measurement matrix with each row being the 1D vector of the selected shifted Hadamard-CS matrix $\overline{\overline{S}}_{p \times q}$; ${\overline{\mathcal{P}}}_{N}$ is the 1D vector form of the 2D image ${\overline{\overline{P}}}_{p \times q}$ and $\overline{n}_L$ is the experimental noise. 

\subsubsection{On the Receiver Side} After the pseudo-random Hadamard transform $\overline{\mathcal{T}}_{L}$ is detected by the receiver, the image can be obtained through the CS reconstruction methods such as LASSO\cite{Emmanuel2008} based on the $\ell_1$ norm.  

\subsection{The DL-CSNet Optimization}
The basic Hadamard-CS image transmission can be effectively optimized by the DL-CSNet, whose functional blocks are shown at the bottom of Fig. \ref{fig:flowchart}. Here, the DL-CSNet is used for three purposes:

\subsubsection{The DL Functionality}
First, the DL-CSNet is used to learn the dictionary basis $\overline{\overline{\mathcal{D}}}_{K_{i+1} \times K_i}^{(i)}$ of the images and their corresponding codes $\overline{\overline{Z}}_{K_{i}}^{(i)}$ through minimizing the two error functions, \ie, the mean square error of the reconstructed image $E_1$ and the mean square error of the reconstructed compressed pseudo-random Hadamard transform $E_2$, as well as the $\ell_1$ norm of the sparse codes $\left|\overline{\overline{Z}}_{K_{i}}^{(i)}\right|$. The loss function of the DL-CSNet $L$ to be minimized is given by
\begin{flalign}\label{eqn:loss}
& L\left(\overline{\overline{\mathcal{D}}}_{K_{i+1} \times K_i}^{(i)}, \overline{Z}_{K_i}^{(i)} \right)  =   \sum_{i=1}^{K} \alpha_i \left|\overline{Z}_{K_i}^{(i)}\right|_1   \\
& +  \beta \left|\left| \overline{\mathcal{P'}}_{N} - \overline{\mathcal{P}}_{N} \right|\right|_2^2  + \gamma \left|\left| \overline{\mathcal{T'}}_{L} - \overline{\mathcal{T}}_{L} \right|\right|_2^2, \nonumber 
\end{flalign}
where $\overline{\overline{\mathcal{D}}}_{K_{i+1} \times K_i}^{(i)}$ is the dictionary basis of the $i$-th fully-connected layer with length $K_i$; the first term is the $\ell_1$ norms of the dictionary codes $\overline{Z}_{K_i}^{(i)}$ of all fully-connected layers; the second term is the mean square error function of the reconstructed image $E_1$; and the third term is the mean square error function of the CS measurements $E_2$; $\alpha_i, \beta, \gamma$ are their corresponding weights.

\subsubsection{The Hadamard Measurement Matrix Optimization Functionality} Then the dictionary basis is used to optimize the Hadamard-CS measurement matrix $\overline{\overline{\mathcal{S}}}_{L \times N}$ through minimizing the mutual coherence of Hadamard measurement matrix \cite{Obermeier-2017}.

\subsubsection{The Lossless Image Reconstruction Functionality}
Finally,  the DL-CSNet reconstructs the lossless image ${\overline{\mathcal{P}}}_{N}$ with the learned dictionary basis as the initial value.
 
\begin{figure*}[th] 
\centerline{\includegraphics[width=1\textwidth]{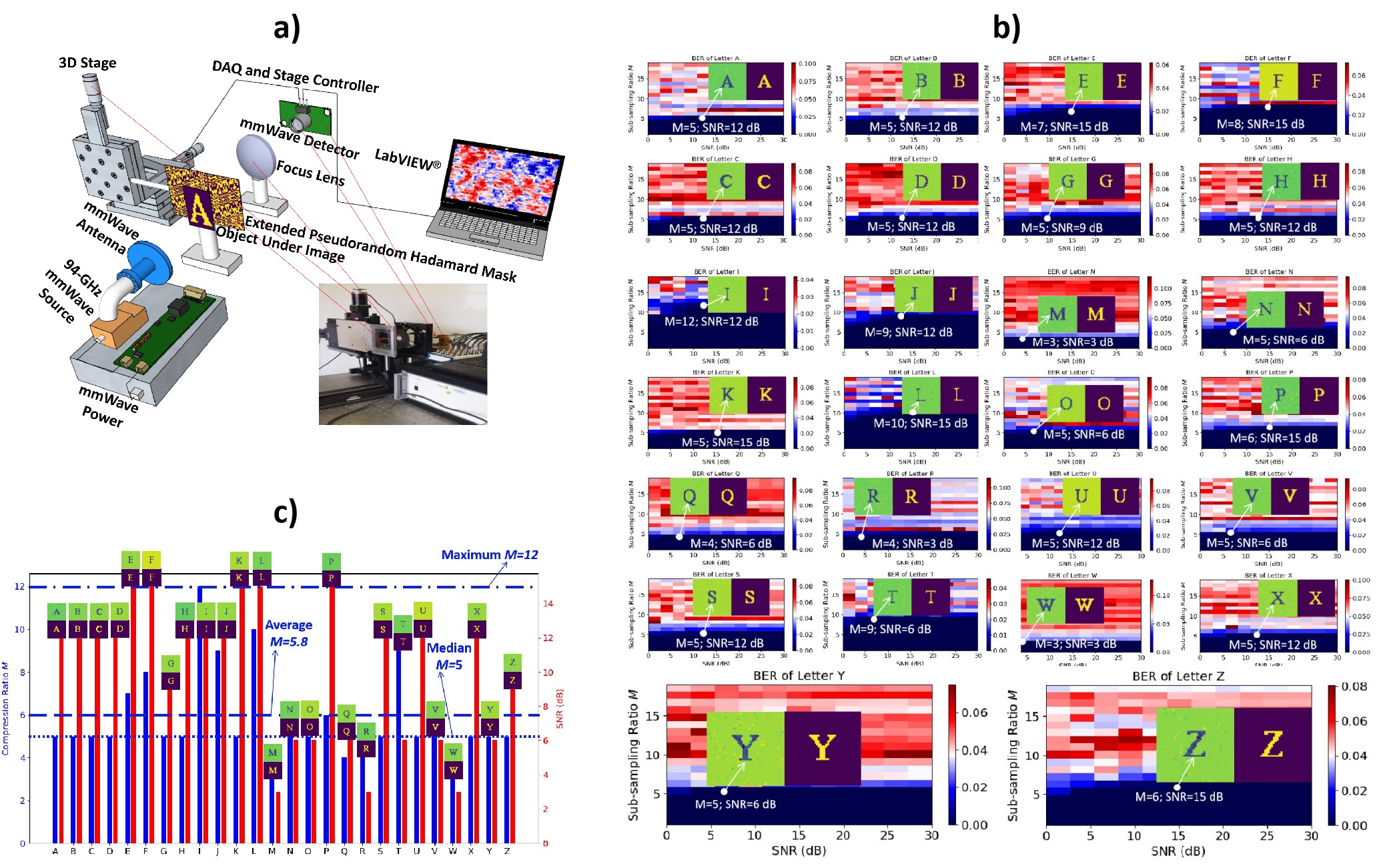}}
\caption{A 94-GHz mmWave Hadamard-CS image transmission prototype: a) experimental setup; b) the DL-CSNet reconstructed letters ``A" to ``Z"; and c) the compression ratio or speed increase and the requried SNR.}
\label{fig:experiment}
\end{figure*}
 
 \section{Experimental Validation}
 The details of the experimental setup of the 94-GHz prototype are shown in Fig \ref{fig:experiment}a). A corrugated circular horn antenna is used to launch the mmWave and is located at the focal point of a 6.8-inch collimating dielectric lens, and the collimated mmWave passes through the image object 10.0-inch away (the letter ``A" here). Besides, immediately adjacent to the image object is the $(p \times q) = (41 \times 43)$ Hadamard-CS mask, with a pixel size of $0.05 \ \hbox{inch} \times 0.05 \ \hbox{inch}$; the $(p \times q) = (41 \times 43)$ Hadamard-CS mask is mounted on a 3D translation stage, which is used to change the optimized Hadamard measurement sequences that are generated by the DL-CSNet. Furthermore, the transmitted field through the pseudo-random Hadamard-CS pattern is focused on the imaging plane of a 1-inch imaging dielectric lens acting as the sum operation, where a corrugated circular horn antenna is connected to a square-law Schottky diode detector to convert the 94 GHz mmWave signal to the base-band signal. A DAQ board is used to collect the signal and the data is transmitted to the LabVIEW\textsuperscript{\textregistered} program running on a computer. After that, the DL-CSNet is used to reconstruct the image from the compressed pseudo-random Hadamard transform. What's more, the image objects used for the mmWave pseudo-random Hadamard transform imaging are letters ``A" to ``Z".

Fig. \ref{fig:experiment}b) shows image reconstruction through the DL-CSNet and the BER for various compression ratios and pixel-equivalent SNRs. It can be seen that a compression ratio up to 12 can be obtained for letter ``I" with the pixel-equivalent power SNR = 12 dB. Also, no image loss occurs after the binary image thresholding. It can be seen that letter ``I" has the maximum $M =12$ with the pixel-equivalent power SNR of 12 dB. The average and median compression ratios are $M =5.8$ and $M =5$ respectively. Finally, Fig. \ref{fig:experiment}b) summarizes the compression ratios and the required SNR for all letters.

\section{Conclusion}
The DL-CSNet AI technique has been developed for high-speed mmWave 5G/6G image transmission. The DL-CSNet can jointly optimize three functionalities, \ie, to learn the dictionary basis of the images for transmission, to optimize the Hadamard measurement matrix, and to reconstruct the lossless images under noisy experimental environment. A 94-GHz prototype has been built and experiment with letters ``A" to ``Z" shows that an image speed increase of 12 has been demonstrated for letter ``I" with a SNR of 12 dB.



\end{document}